\documentclass[prl,showpacs,showkeys,amsfonts,amsmath,twocolumn,superscriptaddress,a4paper,footinbib]{revtex4}
\usepackage{array,amsmath,amsfonts,amssymb,dsfont}
\usepackage[T1]{fontenc}
\usepackage{ae}
\usepackage{times,mathptm}
\usepackage{graphicx}

\DeclareMathOperator{\Tr}{Tr}

\begin{document}
\renewcommand{\i}{{\mathrm i}}
\def\1{\mathchoice{\rm 1\mskip-4.2mu l}{\rm 1\mskip-4.2mu l}{\rm 1\mskip-4.6mu l}{\rm 1\mskip-5.2mu l}}
\newcommand{\ket}[1]{|#1\rangle}
\newcommand{\bra}[1]{\langle #1|}
\newcommand{\braket}[2]{\langle #1|#2\rangle}
\newcommand{\ketbra}[2]{|#1\rangle\langle#2|}
\newcommand{\opelem}[3]{\langle #1|#2|#3\rangle}
\newcommand{\projection}[1]{|#1\rangle\langle#1|}
\newcommand{\scalar}[1]{\langle #1|#1\rangle}
\newcommand{\op}[1]{\mathit{#1}}
\newcommand{\vect}[1]{\boldsymbol{#1}}

\title{Experimental demonstration of the stability of Berry's phase
  for a spin-1/2 particle}

\affiliation{Atominstitut der {\"{O}}sterreichischen Universit{\"{a}}ten,
Stadionallee 2, A-1020 Vienna, Austria}
\affiliation{Department of Physics, ETH Zurich, CH-8093 Z{\"{u}}rich, Switzerland}
\affiliation{Institut Laue Langevin, Bo\^ite Postale 156, F-38042 Grenoble
Cedex 9, France}
\affiliation{Physikalisches Institut, Philosophenweg 12, 69120
  Heidelberg, Germany}
\author{S.~Filipp}
\email{filipp@phys.ethz.ch}
\affiliation{Atominstitut der {\"{O}}sterreichischen Universit{\"{a}}ten,
Stadionallee 2, A-1020 Vienna, Austria}
\affiliation{Department of Physics, ETH Zurich, CH-8093 Z{\"{u}}rich, Switzerland}
\affiliation{Institut Laue Langevin, Bo\^ite Postale 156, F-38042 Grenoble
Cedex 9, France}
\author{J.~Klepp}
\affiliation{Atominstitut der {\"{O}}sterreichischen Universit{\"{a}}ten,
Stadionallee 2, A-1020 Vienna, Austria}
\affiliation{Institut Laue Langevin, Bo\^ite Postale 156, F-38042 Grenoble
Cedex 9, France}
\author{Y.~Hasegawa}
\affiliation{Atominstitut der {\"{O}}sterreichischen Universit{\"{a}}ten,
Stadionallee 2, A-1020 Vienna, Austria}
\author{C.~Plonka-Spehr}
\affiliation{Institut Laue Langevin, Bo\^ite Postale 156, F-38042 Grenoble
Cedex 9, France}
\author{U.~Schmidt}
\affiliation{Physikalisches Institut, Philosophenweg 12, 69120
  Heidelberg, Germany}
\author{P.~Geltenbort}
\affiliation{Institut Laue Langevin, Bo\^ite Postale 156, F-38042 Grenoble
Cedex 9, France}
\author{H.~Rauch}
\affiliation{Atominstitut der {\"{O}}sterreichischen Universit{\"{a}}ten,
Stadionallee 2, A-1020 Vienna, Austria}

\begin{abstract}
The geometric phase has been proposed as a candidate for noise
resilient coherent manipulation of
fragile quantum systems. Since it is determined
 only by the path of the quantum state, the presence of noise fluctuations affects  
the geometric phase in a different way than the dynamical phase. We have experimentally tested the robustness of
 Berry's geometric phase for spin-1/2
particles in a cyclically varying magnetic field.
Using trapped polarized
 ultra-cold neutrons it is
 demonstrated that the geometric
phase contributions to dephasing due to adiabatic field fluctuations vanish for long evolution times. 
\end{abstract}

\pacs{03.65.Vf, 03.65.Yz 03.75.Be} \keywords{geometric
phase; ultra cold neutrons; decoherence}

\maketitle

The rapidly increasing capability to
control and measure quantum states on a single particle level
(see e.~g. \cite{haroche06} and references therein) demands for decoherence free systems and robust manipulation techniques. In order to enable
 high-precision quantum measurements  \cite{roos06}  or quantum
 information processing \cite{nielsen-chuang00} with long coherence
 times, a system should be
 de-coupled from the environment except
 for precisely controllable interactions.
As part of the quest for reliable quantum gates the \emph{geometric phase} has
attracted renewed attention due to its potential resilience against
noise perturbations.

In short, the adiabatic evolution of a
quantum system returning after some time to its initial state gives
rise to an additional phase factor,
termed Berry's phase
\cite{panchberryanandan}. The peculiarity of this phase lies in the
fact that its magnitude is not determined by the dynamics of the system,
i.~e. neither by energy nor by evolution time, but purely by the
evolution path
from the initial to the final state. 
A vast number of
experiments have  verified its characteristics in various systems 
\cite{shapere-wilczek89}. Several extensions, for instance to
non-adiabatic, non-cyclic, non-unitary or non-abelian
geometric phases have been investigated
\cite{aharonov-anandan87,samuel-bhandari88,tong04a,wilczek-zee84}. 
For closed quantum systems the geometric phase is theoretically 
well understood and experimentally verified. However, for open quantum systems the
situation is different in that no general framework has found approval
yet. Concepts of geometric phases for mixed state evolutions
have been introduced theoretically
\cite{sjoqvist00,uhlmann96,ericsson03b} and  inspected
experimentally \cite{du03,ericsson05,du07,klepp08}.
Also dephasing induced by the geometric phase has been
studied theoretically for several settings 
\cite{blais-tremblay03,sarandy-lidar06,whitney,fuentes-guridi05}.
Potential advantages of geometric quantum gates for
quantum information processing have been topic of recent investigation
\cite{zanardi-rasetti99,jones00,falci00}. Furthermore, high fidelity geometric gates are currently used in
ion traps \cite{leibfried03} suggesting Berry's adiabatic
geometric phase as favourable choice for quantum state
manipulations. In \cite{dechiara-palma03} it is shown that the contribution
of the geometric phase to dephasing are  path-dependent like the geometric phase
itself, as experimentally demonstrated in \cite{leek07}, and that they diminish for long evolution
times. 

In this letter we consider the situation of an adiabatic evolution of
a spin-1/2 system and explicitly test the influence of slow
fluctuations onto the Berry phase. We complement the result in \cite{leek07} by analyzing the influence of
evolution time on the geometric dephasing using a dedicated  ultra-cold
neutrons apparatus \cite{filipp08}. We show that the
Berry  phase is robust against adiabatic fluctuations in the
driving field, when the evolution time is longer than the
typical noise correlation time.

Consider a spin-1/2 particle exposed to slowly varying  magnetic
fields. 
The Hamiltonian
\begin{equation}
H(t) = -\mu\vec{\sigma}\cdot \vec{B}(t) = -\mu\left( \vec{\sigma}\cdot \vec{B}_0(t) + \sigma_x
  K(t)\right) 
\label{hamiltonian}
\end{equation}
describes the coupling of a particle 
to the magnetic field $\vec{B}(t)$ by its spin
magnetic moment $\mu$. The magnetic field has magnitude $B(t) \equiv
|\vec{B}(t)|$ and its direction points along the unit vector 
$\vec{n}(t) = (\cos\vartheta(t),\,\sin\vartheta(t) \sin\varphi(t),\,\sin\vartheta(t)
\cos\varphi(t))^T$. $\vec{\sigma} =
(\sigma_x,\sigma_y,\sigma_z)^T$ denote the Pauli matrices and
 $K(t)$ stands for an
additional fluctuating magnetic field along the $x-$axis.
Let $\ket{s_\pm(t)}$ denote the time-dependent spin eigenstates of $H(t)$.
If the system is initially in an eigenstate $\ket{s_\pm(0)}$, it will stay
in an eigenstate during an adiabatic evolution of the B-field. In
other words, the B-field direction and the polarization vector of the
particles' spin
$\vec{s}(t)$ coincide for all
times, where $\vec{s}(t)\equiv \Tr[\vec{\sigma}\cdot\rho(t)]$  for the
system being in the state described by the density matrix $\rho(t)$. In spherical coordinates $\vec{s}(t)$ can be written as 
\begin{equation}
\label{polvector}
\vec{s}(t) = 
 s(\cos\theta(t),\sin\theta(t)\sin\phi(t),\sin\theta(t)\cos\phi(t))^T.
\end{equation}
 Its length $s \equiv |\vec{s}|$ represents the degree of
  polarization. For a pure state $\rho(t) =
  \ket{s_\pm(t)}\bra{s_\pm(t)}$ we have $s= 1$, but in
  general, interactions with the environment lead to mixed states with reduced
  $s<1$ as discussed below. 

Within the adiabatic approximation, where $\theta(t)\approx\vartheta(t)$ and
$\phi(t)\approx\varphi(t)$, a cyclic variation of the B-field coordinates $\vartheta(t)$ and $\varphi(t)$
leads only to a change of the phase of
 the initial state $\ket{s_\pm(0)}$
  after the evolution. The final state
$\ket{s_\pm(T)}=e^{\pm \i(\phi_g +
  \phi_d)}\ket{s_\pm(0)}$ comprises a dynamical ($\phi_d$) and a
geometric ($\phi_g$) phase. $\phi_d= -\int_0^T E(t)dt/\hbar$ is determined by the
integrated instantaneous energies $E(t) = -\mu
\bra{s_\pm(t)}\vec{\sigma}\cdot \vec{B}(t)\ket{s_\pm(t)}$. It depends explicitly
on the dynamics of the state transport. In contrast, the geometric phase
 $\phi_g$ is determined by a surface integral
 proportional to the solid angle $\Omega$ enclosed by the path of the
 state: $\phi_g = -\Omega/2$ for a spin-1/2 particle. It is therefore independent of
energy and time. In particular, the 
B-field we have used in our experiment traces out a path with constant polar angle
$\vartheta$ and varying $\varphi(t) \in [0,2\pi]$.
Without fluctuations ($K(t)=0$) the
geometric phase evaluates in this case to $\phi_g^0 =
-\pi(1-\cos\vartheta)$. 

Field fluctuations in $x$-direction during the evolution are represented by the term $\sigma_x K(t)$ (with equivalent results
holding also for isotropic noise involving $\sigma_{z,y}$ terms
\cite{dechiara-palma03}). Fluctuations in the Larmor frequency
$\omega_L = 2\mu B/\hbar$ are then given by $2\mu
K(t)/\hbar$, which denotes a Gaussian and Markovian noise process with intensity $\sigma_P^2$ and
correlation time $1/\Gamma$, i.~e.~noise bandwidth $\Gamma$.
We assume an upper cut-off frequency of the
noise $\Gamma_{\rm{max}}\ll \omega_L$  such that the field
fluctuations are adiabatic with respect to the Larmor frequency. Later, in the experiment, this is achieved
by adding adequately designed noise to the
field. Consequently, the variations in the path,
and therefore in $\Omega$,
lead to variations in the geometric phase. 
 The random geometric phase
 $\phi_g$
is Gaussian distributed with mean value equal to the
 noise-free case, $\langle\phi_g\rangle = \phi_g^0$. Its variance
 $\sigma^2_g(T)$ depends on the evolution time T and is given by \cite{dechiara-palma03}
\begin{equation}
  \label{geomphasevar}
  \sigma^2_{\phi\,g}(T) = 2 \sigma_P^2\left(\frac{\pi\sin^2\vartheta}{T \omega_L}\right)^2
    \left[\frac{\Gamma T-1+e^{-\Gamma T}}{\Gamma^2}\right].
\end{equation}
The dependence
on the factor $\sin^2\vartheta$ has been tested in
\cite{leek07}. A further intriguing property is that for evolutions,
which are slow relative to the noise fluctuations ($T\gg 1/\Gamma$), the
variance of the 
geometric phase drops to zero for long evolution times as 
the expression in Eq.~(\ref{geomphasevar}) reduces to
$\sigma^2_{\phi\,g}(T)\propto \frac{1}{T}$ \footnote{Note, however, that a
more general theory predicts a shift of the
mean geometric phase and a finite time-scale of the decrease
\cite{whitney,sarandy-lidar06,filipp08-numerical}.}.
This contrasts the behaviour of the variance of the dynamical phase
that increases linearly in time.

\begin{figure}
\centering
\includegraphics[width=40mm]{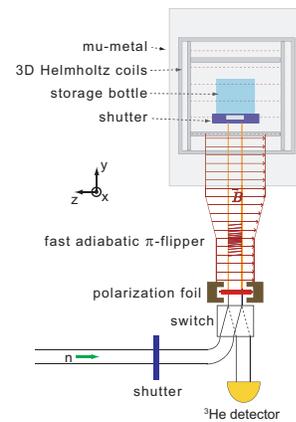}
\caption{Experimental setup: Neutrons passing through 
   polarization foils are guided to the storage bottle, where they are stored for
  typically 10 seconds before letting them fall down to the
  detector. During storage their spin is 
  manipulated by magnetic fields produced by the Helmholtz coils. }
\label{setup}
\end{figure}

In our experiment we have used neutrons as a precisely manipulable
spin-1/2 quantum system.  Exposure to a magnetic field leads to Zeeman energy
splitting of $2 \mu_n B$ with $\mu_n = -9.66\times
10^{-27}~\rm{JT}^{-1}$. The experimental setup is shown in Fig.~\ref{setup} and more details can be found in \cite{filipp08}. Neutrons are
guided from the ultra-cold neutron source at the ILL high flux reactor
through magnetized Fe polarization foils, which give a degree of
polarization of about $90\%$, to the storage
bottle (filling stage). Their low
kinetic energy prohibits penetration through the walls of the bottle. During storage  the spin orientation
of the dilute ($\approx
1~\rm{neutron}/\rm{cm}^{3}$) gas of practically non-interacting spin-1/2
particles can be arbitrarily manipulated by magnetic fields produced by a
3D Helmholtz-coil setup (manipulation stage). The resulting spin polarization is subsequently
analyzed by a combination of a fast adiabatic $\pi-$ flipper ($\approx 99\%$
efficiency) and the
polarization foils before hitting the detector (emptying stage). A full
storage cycle of filling, manipulating and emptying lasts about 70 seconds.

In the following we focus on the 'manipulation stage': First, we have
measured the adiabaticity of the transport of the
polarization vector.  We have compared the initial to the final
polarization after a cyclic variation of the magnetic guide field
($\vec{B}(0)=\vec{B}(T)$) with the neutron spin 
initially
aligned with the static magnetic guide field in the negative
$z-$direction. In order to find identical initial and final
polarization the adiabaticity
condition requires the rate of change of the B-field $|(d\vect{B}/dt)/B|$  to be much
smaller than the Larmor frequency $\omega_L$ of the system. Within the accuracy of
the experiment this applies for a typical rate of change less than
approximately  $0.2\omega_L$ setting an upper limit for the following
measurements.

\begin{figure}
\centering
\includegraphics[width=80mm]{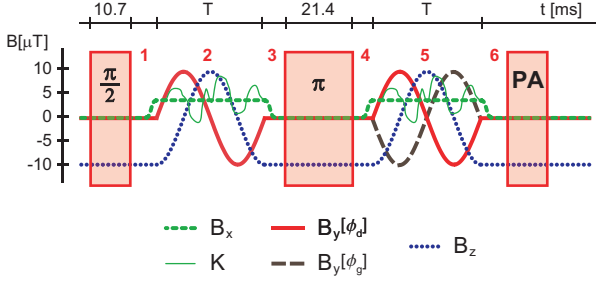}
\caption{Pulses for the B-field ($x-$, $y-$, and $z-$direction) for measuring the geometric
  phase: A $\pi/2$
  pulse  generates a superposition state. This is followed by
  a cyclic evolution with constant B-field magnitude $|B(t)|$. A
  subsequent $\pi$-pulse flips the spin
  such that the preceding unitary evolution can be compensated
  ($\rm{B}_y[\phi_d]$). The resulting spin state is determined by
  the polarization analysis (PA).  
   Measuring the geometric
  phase $\phi_g$ involves a change of the rotation direction while keeping the
  magnitude fixed ($\rm{B}_y[\phi_g]$). Identical fluctuations are generated
  in $x$-direction ($\rm{K}$) for measuring geometric dephasing.}
\label{pulsescheme}
\end{figure}

Secondly, a Ramsey-type interferometric
scheme  similar
to \cite{richardson88} has been employed for the measurement of
Berry's phase. The pulse sequence of the B-field during the manipution stage is
shown in Fig.~\ref{pulsescheme}: The actual evolution is preceded by a 
$\pi/2$-rotation induced by a rf-field in $x-$direction with amplitude 
$1.6~\mu\rm{T}$, duration $10.7~\rm{ms}$ and a frequency resonant with the magnetic guide field $B_z(0) =
-10~\mu\rm{T}$. Starting from the
eigenstate $\ket{s_+(0)}$ this generates an equal
superposition of spin-up and spin-down states,
$\ket{\psi(0)}=\left(\ket{s_+(0)}+\ket{s_-(0)}\right)/\sqrt{2}$.
 A subsequent
adiabatic and cyclic B-field
evolution of duration $T$ induces a relative phase between the
states of $\phi(T)=\phi_g + \phi_d$.
The resulting spin polarization $\vec{s}(T)$ can be analyzed by
 $\pi/2$-pulses, which are offset in phase by zero or $90^\circ$ relative
 to the preparatory $\pi/2$-pulse, thus yielding a rotation of $\vec{s}(T)$
 about the $x-$ or $y-$axis, respectively. A further $\pi$-flip can be induced with high
 efficiency by the
subsequent fast adiabatic $\pi$-flipper. Together with the final projective
measurement along the positive $z$-direction this gives a complete set of measurements of the $\pm x$, $\pm y$ and $\pm
z$  polarization components. In this way the final spin state is
characterized with an efficiency close to
$100\%$.
 The initial degree of polarization  $s_0\equiv|\vec{s}(0)|$ is typically
$75\%$. During the evolution $s_0$ is reduced mainly due to static field  
inhomogeneities across the storage volume -- even without
temporal fluctuations, i.~e. for $K(t)=0$ \cite{filipp08}. Local variations in the
B-field magnitude cause variations in the Zeeman-energy
splitting. Consequently, the relative phases in the final spin superposition states
of the individual neutrons are  randomly distributed, which leads --
on average -- to
 dephasing. This dephasing mechanism
 causes a further exponential polarization loss, $s(T)=s_0
 e^{-T/T_2}$. $T_2=847(40)~\rm{ms}$ has
 been measured by a polarization analysis after
free precession of the spin superposition state in a $10~\mu\rm{T}$
magnetic field, which sufficiently exceeds typical evolution times of $500~\rm{ms}$.

\begin{figure}
  \centering
  \includegraphics[width=80mm]{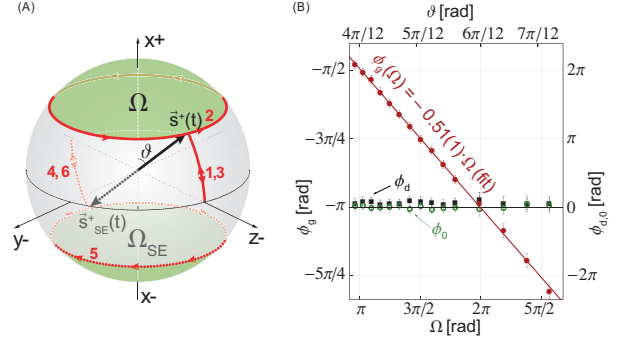}
  \caption{
(A) Path traced out by the spin-up state
    $\vec{s}^+(t)$ on the Bloch-Sphere while following the adiabatic
    changes in the magnetic field.  After
    the first
    cycle (path 1-2-3) the state vector is flipped ($\vec{s}^+\rightarrow
    \vec{s}_{SE}^+$) and traces out the path 4-5-6 on the lower
    hemisphere for the 
    second, echo cycle. The enclosed solid angle $\Omega$ determines
    the geometric phase.
(B) Measured geometric phase $\phi_g$ (filled circles). If the sense of rotation is not reversed in the echo
    pulse, all accumulated phases cancel apart from remaining dynamical
    contributions $\phi_d$ (solid rectangles). Ideally this yields
    the same phase as if there was no B-field evolution at all
    ($\phi_0$ - open circles).}
  \label{geomphase}
\end{figure}

 To measure the geometric phase $\phi_g^0=-\Omega/2$ for $K(t)=0$ the
 magnetic guide field pointing initially in the negative $z$-direction is rotated about the $x$-axis,
i.~e.  $B_{y}(t)=-B_z(0)\sin(\omega\,t)$ and $B_{z}=B_z(0)
\cos(\omega\,t)$ with constant $|\vec{B}(t)|$ (see
Fig.~\ref{pulsescheme}). An additional offset field $B_x$ in
$x$-direction generates a conical section traced out by the magnetic field
vector and -- in the adiabatic limit -- also by the  spin polarization
vector. The enclosed solid angle $\Omega = \pi(1-\cos\vartheta)$ is determined by the cone angle $\vartheta =
\tan^{-1} B_z/B_{x}$.
To eliminate contributions from the dynamical phase $\phi_d$ we invoke a
spin-echo scheme \cite{abragam61,bertlmann04}. The according evolution
path of the spin-up component
$\vec{s}^+(t)\equiv\Tr\left[\vec{\sigma}\ket{s_+(t)}\bra{s_+(t)}\right]$
of the superposition state on the Bloch-sphere  
 is illustrated  in
Fig.~\ref{geomphase}(A). Depending on the rotation
    direction the solid angle enclosed by the path on the lower
    hemisphere $\Omega_{SE} = \pm\Omega$. Thus, if the direction of
rotation is reversed after a $\pi$-pulse and
the field amplitude is kept constant, the geometric phase doubles -- due to its dependence on the directed solid
  angle -- while the dynamical
phase cancels. 
Both the accumulation of the
geometric phase and the cancellation of the dynamical phase has been
measured using the pulse sequence drawn in Fig.~\ref{pulsescheme}
for $T = 200~\rm{ms}$. The polar angle $\vartheta$ and consequently
the solid angle $\Omega$ is varied by
choosing different $B_x$ offset fields. The ratio
$\omega/\omega_L =2\pi/(T\omega_L)\approx 0.017$ ensures adiabaticity of the evolution.
In
Fig.~\ref{geomphase}(B) the measured geometric phase $\phi_g$ is
plotted as a function of the 
solid angle $\Omega$.  The fit to the measured data
yields $\phi_g^0 = -0.51(1) \Omega$ which is in good agreement with the
expected $\phi_g^0 = -\Omega/2$. Residual dynamical phase contributions, which are
not compensated by the spin echo, are measured to be
$0.22~\rm{rad}$. These are determined by the phase difference in the
final polarization between the spin-echo 
 with identical evolution (Fig.~\ref{pulsescheme} for $B_y[\phi_d]$)
 and without
evolution at all ($B_z(t)=\rm{const.}$, $B_x(t)=B_y(t)=0$).

\begin{figure}[htbp]
  \centering
  \includegraphics[width=70mm]{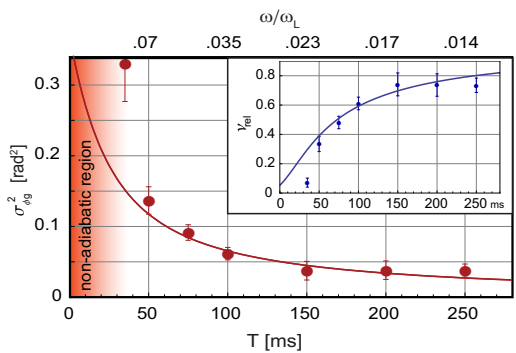}
  \caption{The variance $\sigma_{\phi\,g}^2(T)$ of the geometric phase as a
    function of the evolution time $T$ in a fluctuating magnetic
    field. The variance decreases for longer evolution
    times following closely the theoretical prediction in 
    Eq.~(\ref{geomphasevar}) (solid line). The
    inset shows the increase of the degree of polarization $\nu_{\rm{rel}}$
    relative to
    the noise-free evolution along with the theoretical predictions. 
}
  \label{variances}
\end{figure}

For testing the stability  of the geometric phase we
generate field fluctuations $K(t)$ (c.~f. Eq. (\ref{hamiltonian})) with Lorentzian power spectrum \cite{filipp08-numerical}, a bandwidth of
$\Gamma=100~\rm{rad\,s}^{-1}$ and intensity  $\sigma_P^2 =
4~\mu\rm{T}^2$ as the mean square deviation of the fluctuations.  We applied a smooth rectangular-shaped
 window function to the noise in the time-domain to avoid non-adiabatic and non-cyclic
 effects.  To test the time dependence
of the variance of the geometric phase $\sigma_{\phi\,g}^2(T)$
given by Eq.~(\ref{geomphasevar}) the evolution time $T$ is
changed from $T = 35~\rm{ms}$ to $T= 250~\rm{ms}$. The different fluctuations in subsequent storage cycles lead to different phases
$\phi$ of the final state. But since the
noise is identical for both first and second part of the spin-echo,
these difference can only originate in the geometric phase. The
dynamical phase cancels as before in the fluctuation-free
measurements. The average over several storage cycles, i.~e. several
noise patterns, gives a further shrinking of
the length of the polarization vector $s(T)$ of the final state
additional to the unavoidable polarization losses discussed above: In fact, for Gaussian
  distributed $\phi$  we obtain  $\langle \cos \phi\rangle =
  \exp[-\sigma_\phi^2/2]\cos\langle \phi\rangle$ and same for  $\langle \sin \phi\rangle$
in Eq.~(\ref{polvector}). 
Consequently, the purely
  geometric dephasing gives $s_n(T) =
s(T)\exp\left[-\left(4\sigma_{\phi\,g}(T)\right)^2/2\right]$, where
the factor $4$ is
  due to the particular type of measurement \footnote{Both the preparation of
  a superposition state and the spin-echo gives a factor of two for
  the phase.} and small fluctuations in $\theta$ are neglected. 
To separate the unavoidable polarization losses from the geometric dephasing
  the geometric phase has been measured with and
without fluctuations and the ratio of the corresponding degrees
of polarization $\nu_{\rm{rel}}(T) \equiv s_n(T)/s(T) =
\exp[-\left(4\sigma_{\phi\,g}(T)\right)^2/2]$
gives the variance of the
geometric phase $\sigma_{\phi\,g}^2$. 300  different noise
realizations have been performed for each value of T, where a
sequence of
six storage cycles forms the polarization analysis.
In Fig.~\ref{variances} we have plotted the decrease of $\sigma_{\phi\,g}^2$ as a function
of the evolution time $T$ and fixed noise-free geometric phase
$\phi_g^0 = -2.56\,\rm{rad}$. The inset shows the corresponding increase in the relative
degree of polarization $\nu_{\rm{rel}}$. Error bars stem from the
limited number of noise realizations.
  The solid line indicates
the theoretical predictions given by Eq.~(\ref{geomphasevar}) for the adjusted experimental parameters
without free parameters. Due to the low-pass filtering of the coils and influences from a non-adiabatic manipulation the data-point at $T=35~\rm{ms}$ deviates from the
theory curve by $3\,\sigma$.  We have also verified that
the mean geomtric phase
remains unaffected: $\Delta=|\langle \bar{\phi}_g \rangle -
\phi_g^0| = 0\pm0.1~\rm{rad}$, where $\bar{\phi}_g$ denotes
the average of the geometric phase averaged over the different values
of T.

In summary, we have measured the stability of the adiabatic
geometric phase with respect to fluctuations in the parameters of the driving Hamiltonian as a function of
evolution time. A spin-echo technique allowed for the observation of
the purely geometric part of the dephasing of the quantum state by
eliminating dynamical contributions. The acquired data shows very good agreement with 
theoretical predictions and  demonstrate the vanishing influence of
geometric dephasing for slow evolutions. Clearly, when considering
quantum gates a compromise has to be found between the superior noise
resilience but slower execution speed compared
to dynamical phase gates. In this context generalized
settings involving non-adiabatic geometric phases \cite{zhu-zanardi05}
provide an interesting perspective for future experimental efforts. In
the adiabatic regime the results presented above
demonstrate that 
the geometric phase can indeed be useful for high-fidelity quantum
state manipulations. 

This work was supported by the Austrian Science
Foundation (FWF)  and
the Japan Science and Technology Agency (JST). The authors want to thank
 K.~Durstberger, E.~Sj\"oqvist, S.~Sponar, and R.~Whitney, as well as R.~Loidl, P.~Pataki, E.~Tischler, and
T.~Brenner for
valuable discussions and for technical assistance.


\begin{thebibliography}{30}
\expandafter\ifx\csname natexlab\endcsname\relax\def\natexlab#1{#1}\fi
\expandafter\ifx\csname bibnamefont\endcsname\relax
  \def\bibnamefont#1{#1}\fi
\expandafter\ifx\csname bibfnamefont\endcsname\relax
  \def\bibfnamefont#1{#1}\fi
\expandafter\ifx\csname citenamefont\endcsname\relax
  \def\citenamefont#1{#1}\fi
\expandafter\ifx\csname url\endcsname\relax
  \def\url#1{\texttt{#1}}\fi
\expandafter\ifx\csname urlprefix\endcsname\relax\def\urlprefix{URL }\fi
\providecommand{\bibinfo}[2]{#2}
\providecommand{\eprint}[2][]{\url{#2}}

\bibitem[{\citenamefont{Haroche and Raimond}(2006)}]{haroche06}
\bibinfo{author}{\bibfnamefont{S.}~\bibnamefont{Haroche}} \bibnamefont{and}
  \bibinfo{author}{\bibfnamefont{J.-M.} \bibnamefont{Raimond}},
  \emph{\bibinfo{title}{Exploring the {Q}uantum}} (\bibinfo{publisher}{Oxford
  Univ. Press}, \bibinfo{year}{2006}).


\bibitem[{\citenamefont{Roos et~al.}(2006)\citenamefont{Roos, Chwalla, Kim,
  Riebe, and Blatt}}]{roos06}
\bibinfo{author}{\bibfnamefont{C.~F.} \bibnamefont{Roos} {\em et al.}},
  \bibinfo{journal}{Nature} \textbf{\bibinfo{volume}{443}},
  \bibinfo{pages}{316} (\bibinfo{year}{2006}).

\bibitem[{\citenamefont{Nielsen and Chuang}(2000)}]{nielsen-chuang00}
\bibinfo{author}{\bibfnamefont{M.~A.} \bibnamefont{Nielsen}} \bibnamefont{and}
  \bibinfo{author}{\bibfnamefont{I.~L.} \bibnamefont{Chuang}},
  \emph{\bibinfo{title}{Quantum Computation and Quantum Information}}
  (\bibinfo{publisher}{Cambridge University Press},
  \bibinfo{address}{Cambridge, U. K.}, \bibinfo{year}{2000}).


\bibitem[{\citenamefont{Pancharatnam}(1956)}]{panchberryanandan}
\bibinfo{author}{\bibfnamefont{S.}~\bibnamefont{Pancharatnam}},
  \bibinfo{journal}{Proc. Ind. Acad. Sci.} \textbf{\bibinfo{volume}{A44}},
  \bibinfo{pages}{247} (\bibinfo{year}{1956});
\bibinfo{author}{\bibfnamefont{M.~V.} \bibnamefont{Berry}},
  \bibinfo{journal}{Proc. Roy. Soc. Lond. A} \textbf{\bibinfo{volume}{392}},
  \bibinfo{pages}{45} (\bibinfo{year}{1984});
\bibinfo{author}{\bibfnamefont{J.}~\bibnamefont{Anandan}},
  \bibinfo{journal}{Nature} \textbf{\bibinfo{volume}{360}},
  \bibinfo{pages}{307} (\bibinfo{year}{1992}).


\bibitem[{\citenamefont{Shapere and Wilczek}(1989)}]{shapere-wilczek89}
\bibinfo{editor}{\bibfnamefont{A.}~\bibnamefont{Shapere}} \bibnamefont{and}
  \bibinfo{editor}{\bibfnamefont{F.}~\bibnamefont{Wilczek}}, eds.,
  \emph{\bibinfo{title}{Geometric Phases in Physics}}
  (\bibinfo{publisher}{World Scientific}, \bibinfo{address}{Singapore},
  \bibinfo{year}{1989}).

\bibitem[{\citenamefont{Aharonov and Anandan}(1987)}]{aharonov-anandan87}
\bibinfo{author}{\bibfnamefont{Y.}~\bibnamefont{Aharonov}} \bibnamefont{and}
  \bibinfo{author}{\bibfnamefont{J.} \bibnamefont{Anandan}},
  \bibinfo{journal}{Phys. Rev. Lett} \textbf{\bibinfo{volume}{58}},
  \bibinfo{pages}{1593} (\bibinfo{year}{1987}).

\bibitem[{\citenamefont{Samuel and Bhandari}(1988)}]{samuel-bhandari88}
\bibinfo{author}{\bibfnamefont{J.}~\bibnamefont{Samuel}} \bibnamefont{and}
  \bibinfo{author}{\bibfnamefont{R.}~\bibnamefont{Bhandari}},
  \bibinfo{journal}{Phys. Rev. Lett.} \textbf{\bibinfo{volume}{60}},
  \bibinfo{pages}{2339} (\bibinfo{year}{1988}).

\bibitem[{\citenamefont{Tong et~al.}(2004)\citenamefont{Tong, Sj{\"o}qvist,
  Kwek, and Oh}}]{tong04a}
\bibinfo{author}{\bibfnamefont{D.~M.} \bibnamefont{Tong}},
  \bibinfo{author}{\bibfnamefont{E.}~\bibnamefont{Sj{\"o}qvist}},
  \bibinfo{author}{\bibfnamefont{L.~C.} \bibnamefont{Kwek}}, \bibnamefont{and}
  \bibinfo{author}{\bibfnamefont{C.~H.} \bibnamefont{Oh}},
  \bibinfo{journal}{Phys. Rev. Lett.} \textbf{\bibinfo{volume}{93}},
  \bibinfo{pages}{080405} (\bibinfo{year}{2004}).

\bibitem[{\citenamefont{Wilczek and Zee}(1984)}]{wilczek-zee84}
\bibinfo{author}{\bibfnamefont{F.}~\bibnamefont{Wilczek}} \bibnamefont{and}
  \bibinfo{author}{\bibfnamefont{A.}~\bibnamefont{Zee}},
  \bibinfo{journal}{Phys. Rev. Lett.} \textbf{\bibinfo{volume}{52}},
  \bibinfo{pages}{2111} (\bibinfo{year}{1984}).


\bibitem[{\citenamefont{Sj{\"o}qvist et~al.}(2000)\citenamefont{Sj{\"o}qvist,
  Pati, Ekert, Anandan, Ericsson, Oi, and Vedral}}]{sjoqvist00}
\bibinfo{author}{\bibfnamefont{E.}~\bibnamefont{Sj{\"o}qvist} {\em et al.}},
  \bibinfo{journal}{Phys. Rev. Lett.} \textbf{\bibinfo{volume}{85}},
  \bibinfo{pages}{2845} (\bibinfo{year}{2000}).

\bibitem[{\citenamefont{Uhlmann}(1996)}]{uhlmann96}
\bibinfo{author}{\bibfnamefont{A.}~\bibnamefont{Uhlmann}}, \bibinfo{journal}{J.
  Geom. Phys.} \textbf{\bibinfo{volume}{18}}, \bibinfo{pages}{76}
  (\bibinfo{year}{1996}).

\bibitem[{\citenamefont{Ericsson et~al.}(2003)\citenamefont{Ericsson, Pati,
  Sj{\"o}qvist, Br\"{a}nnlund, and Oi}}]{ericsson03b}
\bibinfo{author}{\bibfnamefont{M.}~\bibnamefont{Ericsson} {\em et al.}},
  \bibinfo{journal}{Phys. Rev. Lett.} \textbf{\bibinfo{volume}{91}},
  \bibinfo{pages}{090405} (\bibinfo{year}{2003}).



 \bibitem[{\citenamefont{Du et~al.}(2003)\citenamefont{Du, Zou, Shi, Kwek, Pan,
   Oh, Ekert, Oi, and Ericsson}}]{du03}
 \bibinfo{author}{\bibfnamefont{J.}~\bibnamefont{Du} {\em et al.}},
   \bibinfo{journal}{Phys. Rev. Lett.} \textbf{\bibinfo{volume}{91}},
   \bibinfo{pages}{100403} (\bibinfo{year}{2003}).

\bibitem[{\citenamefont{Ericsson et~al.}(2005)\citenamefont{Ericsson, Achilles,
  Barreiro, Branning, Peters, and Kwiat}}]{ericsson05}
\bibinfo{author}{\bibfnamefont{M.}~\bibnamefont{Ericsson} {\em et al.}},
  \bibinfo{journal}{Phys. Rev. Lett.} \textbf{\bibinfo{volume}{94}},
  \bibinfo{pages}{050401} (\bibinfo{year}{2005}).


\bibitem[{\citenamefont{Du et~al.}(2007)\citenamefont{Du, Shi, Zhu, Vedral,
  Peng, and Suter}}]{du07}
\bibinfo{author}{\bibfnamefont{J.}~\bibnamefont{Du} {\em et al.}},
 \bibinfo{howpublished}{arXiv:0710.5804v1
  [quant-ph]} (\bibinfo{year}{2007}).

\bibitem[{\citenamefont{Klepp et~al.}()\citenamefont{Klepp, Sponar, Filipp,
  Lettner, Badurek, and Hasegawa}}]{klepp08}
\bibinfo{author}{\bibfnamefont{J.}~\bibnamefont{Klepp} {\em et al.}},
\bibinfo{journal}{Phys. Rev. Lett.} \textbf{\bibinfo{volume}{101}},
  \bibinfo{pages}{015404} (\bibinfo{year}{2008}).

\bibitem[{\citenamefont{Blais and Tremblay}(2003)}]{blais-tremblay03}
\bibinfo{author}{\bibfnamefont{A.}~\bibnamefont{Blais}} \bibnamefont{and}
  \bibinfo{author}{\bibfnamefont{A.}~\bibfnamefont{M.}~\bibfnamefont{S.}~\bibnamefont{Tremblay}},
  \bibinfo{journal}{Phys. Rev. A} \textbf{\bibinfo{volume}{67}},
  \bibinfo{pages}{012308} (\bibinfo{year}{2003}).

\bibitem[{\citenamefont{Sarandy and Lidar}(2006)}]{sarandy-lidar06}
\bibinfo{author}{\bibfnamefont{M.~S.} \bibnamefont{Sarandy}} \bibnamefont{and}
  \bibinfo{author}{\bibfnamefont{D.~A.} \bibnamefont{Lidar}},
  \bibinfo{journal}{Phys. Rev. A} \textbf{\bibinfo{volume}{73}},
  \bibinfo{pages}{062101} (\bibinfo{year}{2006}).




 \bibitem[{\citenamefont{Whitney et~al.}(2005)\citenamefont{Whitney, Makhlin,
   Shnirman, and Gefen}}]{whitney}
 \bibinfo{author}{\bibfnamefont{R.~S.} \bibnamefont{Whitney}},
   \bibinfo{author}{\bibfnamefont{Y.}~\bibnamefont{Makhlin}},
   \bibinfo{author}{\bibfnamefont{A.}~\bibnamefont{Shnirman}}, \bibnamefont{and}
   \bibinfo{author}{\bibfnamefont{Y.}~\bibnamefont{Gefen}},
   \bibinfo{journal}{Phys. Rev. Lett.} \textbf{\bibinfo{volume}{94}},
   \bibinfo{pages}{070407} (\bibinfo{year}{2005});
 \bibinfo{author}{\bibfnamefont{R.~S.} \bibnamefont{Whitney}} \bibnamefont{and}
   \bibinfo{author}{\bibfnamefont{Y.}~\bibnamefont{Gefen}},
   \bibinfo{journal}{Phys. Rev. Lett.} \textbf{\bibinfo{volume}{90}},
   \bibinfo{pages}{190402} (\bibinfo{year}{2003}).


\bibitem[{\citenamefont{Fuentes-Guridi
  et~al.}(2005)\citenamefont{Fuentes-Guridi, Girelli, and
  Livine}}]{fuentes-guridi05}
\bibinfo{author}{\bibfnamefont{I.}~\bibnamefont{Fuentes-Guridi}},
  \bibinfo{author}{\bibfnamefont{F.}~\bibnamefont{Girelli}}, \bibnamefont{and}
  \bibinfo{author}{\bibfnamefont{E.}~\bibnamefont{Livine}},
  \bibinfo{journal}{Phys. Rev. Lett.} \textbf{\bibinfo{volume}{94}},
  \bibinfo{pages}{020503} (\bibinfo{year}{2005}).

\bibitem[{\citenamefont{Zanardi and Rasetti}(1999)}]{zanardi-rasetti99}
\bibinfo{author}{\bibfnamefont{P.}~\bibnamefont{Zanardi}} \bibnamefont{and}
  \bibinfo{author}{\bibfnamefont{M.}~\bibnamefont{Rasetti}},
  \bibinfo{journal}{Phys. Lett. A} \textbf{\bibinfo{volume}{294}},
  \bibinfo{pages}{94} (\bibinfo{year}{1999}).

\bibitem[{\citenamefont{Jones et~al.}(2000)\citenamefont{Jones, Vedral, Ekert,
  and Castagnoli}}]{jones00}
\bibinfo{author}{\bibfnamefont{J.~A.} \bibnamefont{Jones}},
  \bibinfo{author}{\bibfnamefont{V.}~\bibnamefont{Vedral}},
  \bibinfo{author}{\bibfnamefont{A.}~\bibnamefont{Ekert}}, \bibnamefont{and}
  \bibinfo{author}{\bibfnamefont{G.}~\bibnamefont{Castagnoli}},
  \bibinfo{journal}{Nature} \textbf{\bibinfo{volume}{403}},
  \bibinfo{pages}{869} (\bibinfo{year}{2000}).

\bibitem[{\citenamefont{Falci et~al.}(2000)\citenamefont{Falci, Fazio, Palma,
  Siewert, and Vedral}}]{falci00}
\bibinfo{author}{\bibfnamefont{G.}~\bibnamefont{Falci} {\em et al.}},
  \bibinfo{journal}{Nature} \textbf{\bibinfo{volume}{407}},
  \bibinfo{pages}{335} (\bibinfo{year}{2000}).

\bibitem[{\citenamefont{Leibfried et~al.}(2003)\citenamefont{Leibfried,
   DeMarco, Meyer, Lucas, Barrett, Britton, Itano, Jelenkovi´c, Langer,
   Rosenband et~al.}}]{leibfried03}
 \bibinfo{author}{\bibfnamefont{D.}~\bibnamefont{Leibfried} {\em et al.}},
    \bibinfo{journal}{Nature}
   \textbf{\bibinfo{volume}{422}}, \bibinfo{pages}{412} (\bibinfo{year}{2003}).


\bibitem[{\citenamefont{Chiara and Palma}(2003)}]{dechiara-palma03}
\bibinfo{author}{\bibfnamefont{G.} \bibnamefont{DeChiara}} \bibnamefont{and}
  \bibinfo{author}{\bibfnamefont{G.~M.}~\bibnamefont{Palma}},
  \bibinfo{journal}{Phys. Rev. Lett.} \textbf{\bibinfo{volume}{91}},
  \bibinfo{pages}{090404} (\bibinfo{year}{2003}).

\bibitem[{\citenamefont{Leek et~al.}(2007)\citenamefont{Leek, Fink, Blais,
  Bianchetti, G{\"o}ppl, Schuster, Frunzio, Schoelkopf, and Wallraff}}]{leek07}
\bibinfo{author}{\bibfnamefont{P.~J.} \bibnamefont{Leek} {\em et al.}},
  \bibinfo{journal}{Science} \textbf{\bibinfo{volume}{318}},
  \bibinfo{pages}{1889} (\bibinfo{year}{2007}).




 \bibitem[{\citenamefont{Filipp et~al.}(2008)\citenamefont{Filipp, Hasegawa,
   Klepp, Plonka, Schmidt, Geltenbort, and Rauch}}]{filipp08}
 \bibinfo{author}{\bibfnamefont{S.}~\bibnamefont{Filipp} {\em et al.}},
   \bibinfo{journal}{NIMA} \textbf{\bibinfo{volume}{598}},
   \bibinfo{pages}{571} (\bibinfo{year}{2009}).

\bibitem[{\citenamefont{Richardson et~al.}(1988)\citenamefont{Richardson,
  Kilvington, Green, and Lamoreaux}}]{richardson88}
\bibinfo{author}{\bibfnamefont{D.~J.}~\bibnamefont{Richardson}},
  \bibinfo{author}{\bibfnamefont{A.~I.}~\bibnamefont{Kilvington}},
  \bibinfo{author}{\bibfnamefont{K.}~\bibnamefont{Green}}, \bibnamefont{and}
  \bibinfo{author}{\bibfnamefont{S.~K.}~\bibnamefont{Lamoreaux}},
  \bibinfo{journal}{Phys. Rev. Lett.} \textbf{\bibinfo{volume}{61}},
  \bibinfo{pages}{2030} (\bibinfo{year}{1988}).


\bibitem[{\citenamefont{Abragam}(1961)\citenamefont{Abragam}}]{abragam61}
\bibinfo{author}{\bibfnamefont{A.}~\bibnamefont{Abragam}},
\emph{\bibinfo{title}{Principles of Nuclear Magnetism}}
(\bibinfo{publisher}{Oxford Univ. Press},
\bibinfo{address}{Oxford, U.~K.}, \bibinfo{year}{1961}).

\bibitem[{\citenamefont{Bertlmann et~al.}(2004)\citenamefont{Bertlmann,
  Durstberger, Hasegawa, and Hiesmayr}}]{bertlmann04}
\bibinfo{author}{\bibfnamefont{R.~A.}~\bibnamefont{Bertlmann}},
  \bibinfo{author}{\bibfnamefont{K.}~\bibnamefont{Durstberger}},
  \bibinfo{author}{\bibfnamefont{Y.}~\bibnamefont{Hasegawa}}, \bibnamefont{and}
  \bibinfo{author}{\bibfnamefont{B.~C.}~\bibnamefont{Hiesmayr}},
  \bibinfo{journal}{Phys. Rev. A} \textbf{\bibinfo{volume}{69}},
  \bibinfo{pages}{032112} (\bibinfo{year}{2004}).

\bibitem[{\citenamefont{Filipp}(2008)}]{filipp08-numerical}
\bibinfo{author}{\bibfnamefont{S.}~\bibnamefont{Filipp}},
  \bibinfo{journal}{Eur. Phys. J. ST}
  \textbf{\bibinfo{volume}{160}},
  \bibinfo{pages}{165}  (\bibinfo{year}{2008}).

\bibitem[{\citenamefont{Zhu}(2005)\citenamefont{Zhu, and Zandardi}}]{zhu-zanardi05}
\bibinfo{author}{\bibfnamefont{S.-L.}~\bibnamefont{Zhu}},
\bibinfo{author}{\bibfnamefont{P.}~\bibnamefont{Zanardi}},
  \bibinfo{journal}{Phys. Rev. A} \textbf{\bibinfo{volume}{72}},
  \bibinfo{pages}{020301(R)} (\bibinfo{year}{2005}).

\end{thebibliography}

\end{document}